\documentclass[doublecol]{epl2}
\usepackage{epsfig}
\usepackage{color,soul}
\sethlcolor{red}
\title{
Role of the cooling rate in the stability of the superconducting phase of  (TMTSF)$_2$ClO$_4$
}
\author{
S. Haddad, I. Sfar, S. Charfi-Kaddour \and R. Bennaceur
}
\institute{
Laboratoire de Physique de la Mati\`ere Condens\'ee, D\'epartement de Physique,
Facult\'e des Sciences de Tunis, Campus universitaire 1060 Tunis, Tunisia
}
\date{today}

\abstract{
The noncentrosymmetric ClO$_4$ anions of the organic superconductor (TMTSF)$_2$ClO$_4$ order below 24K. The size of domains where the anions are ordered is substantially dependent on the cooling rate which is a key parameter for the stability of the low temperature electronic ground states.
We study the effect of the cooling rate on the SC phase within a self consistent approach in the framework of the time dependent Ginzburg-Landau theory taking into account the superconducting fluctuations. We derive the superconducting transition temperature which is found to decrease with increasing cooling rate in agreement with recent experimental data.
}

\begin{document}
\maketitle
%
The Bechgaard salts (TMTSF)$_2$X, where TMTSF denotes the organic molecule
tetramethyltetraselenafulvalene and X stands for the inorganic anion (X= PF$_6$, ClO$_4$, ReO$_4$ etc.), continue to be under close scrutiny due to their rich phase diagram exhibiting a variety of ground states such as metallic and superconducting (SC) states as well as spin density wave (SDW) phases\cite{Revue,Revue1}. A key feature of this phase diagram is the proximity of the insulating SDW phase to the superconducting one, which raises up a worthwile question about the interplay between these two states: wether they coexist or compete.\newline
Transport measurements on (TMTSF)$_2$PF$_6$ \cite{Vuletic,Chaikin} revealed the presence of an inhomogeneous superconducting state in the vicinity of the border between the SDW phase and the superconducting one. This feature has been ascribed to the coexistence of macroscopic SDW and superconducting slabs under high pressure.
Contraray to the (TMTSF)$_2$PF$_6$ salt, (TMTSF)$_2$ClO$_4$ undergoes a superconducting phase transition at ambient pressure. The stability of this phase is substantially dependent on the rate at which the sample is cooled through the structural transition temperature T$_{\mathrm{AO}}$. Below this temperature, the noncentrosymmetric ClO$^-_4$ anions order by alternating their orientations along the $b$ direction perpendicular to the stacks of the TMTSF molecules arranged along the $a$ axis.\newline
In the relaxed state, obtained by slowly cooling the sample at a rate less than 0.1K/mn, the ClO$^-_4$ anions ordering occurs at 24K at ambient pressure and superconductivity appears at 1.2K.
However, in the quenched sample, rapidly cooled at a rate of at least 50K/mn, the anion ordering can no more be achieved and the anions are randomly orientated. In this case, the superconducting ground state is destroyed and a SDW phase is established at 6K.\

What is remarkable with the anion ordering (AO) is the possibility to have intermediate states with a mixture of SDW and SC regions. This feature has been first reported by transport \cite{Schwenk} and specific heat \cite{Pesty} measurements at different cooling rates which have been corroborated by X ray diffraction studies \cite{Pouget}. Based on a muon spin rotation study, Greer {\it et al.} have also observed, for intermediate cooling rates, a mixed state with macroscopic sized domains of superconducting and SDW phases\cite{Greer}.\

Moreover, it has been found that, by increasing the cooling rate, the SC transition temperature decreases and ends at 0.7K for a cooling rate of 36K/min according to
Meissner signal measurements \cite{Park,Mat}.\

The cooling rate dependence of the SC transition temperature T$_c$ has been recently re-investigated by Joo {\it et al.} who derived the phase diagram of (TMTSF)$_2$ClO$_4$ for different cooling rates \cite{JooPhD,Joo04}. The results revealed that T$_c$ starts to decrease rapidly for small quenching rates (less than 12K/mn) but the decrease is slowed down as the cooling rate increases and T$_c$ tends to saturate at 0.9K before the suppression of the SC phase at a rate of 14K/mn.\

For the rates from 11K/mn to 14K/mn, the authors reported the coexistence of a SC state and a SDW phase. The latter is revealed by the upturn of the resistivity preceding its drop at the SC transition temperature in accordance with the early measurements of Schwenk {\it et al.}\cite{Schwenk}.\

This behavior has been ascribed to the presence of ordered ClO$^-_4$ anions domains where superconductivity develops and disordered regions where SDW phase appears. These two ground states do not coexist microscopically but are rather segregated as pointed out in Ref.\cite{Schwenk}.\

For slow cooling rates, the size of disordered domains is small in such a way that they can be considered as defects or non-magnetic scattering centers, which are known to induce a sharp decrease of the transition temperature of non-s wave SC phase according to the Abrikosov-Gor'kov theory\cite{Abrikos}. Therefore, the sensitivity of T$_c$ to the cooling rate in the (TMTSF)$_2$ClO$_4$ has been considered as a proof for an unconventional SC state \cite{Mat,Joo04}.\

As the cooling rate increases, the disordered domains can no more act as point defects and the sensitivity of T$_c$ to the quenching is then more and more reduced which explains the slowing down of the rate at which T$_c$  decreases as found by Joo {\it et al.} \cite{JooPhD,Joo04}.\ 

In the quenched state, the SDW islands merge to form larger domains at the expense of the ordered ones. As a result the SC phase is completely suppressed and a pure SDW phase is stabilized.
It is worth to note that a pure SDW (SC phase) has not been observed, there are always tiny ordered (disordered) domains where the SC (SDW) phase develops. \newline\\

A first attempts to interpret the effect of the AO on the ground state of the (TMTSF)$_2$ClO$_4$ salt was undertook by H\'eritier {\it et al.} \cite{Heritier84}
who derived a rough estimate of the SDW transition temperature in the relaxed state and in the quenched one. The authors argued that AO may perturb the nesting properties of the Fermi surface and as a consequence, the SDW phase is destabilized at ambient pressure in the relaxed state.
However, the possibility of a phase mixture has not been addressed.\
In Ref\cite{ISCOM05}, renormalization group calculations were proposed taking into account the anion ordering which is assumed to develop over the whole sample.
The results showed that the singlet SC state is the most stable phase in the relaxed state while the SDW phase will develop in the quenched state. 
For the intermediate cooling rates, the results revealed a microscopically coexistence between the SC and the SDW phases while experiments suggest that these phases are rather segregated as discussed above.\
To deal with this phase segregation, a somewhat different renormalization group (RG) approach has been proposed based on a structure of ordered and disordered anion domains. According to this model, the segregated phase is expected to be the unique possible description of the intermediate state\cite{Noomen}. However, these RG calculations are carried out in the one dimensional regime far from the low temperature ordered states. \newline
As far as we know, no theoretical study of these states, taking into account AO, has yet been provided, in particular for the intermediate cooling rate.\

In this letter, we propose a model to describe the dependence of the SC transition temperature on the cooling rate within a selfconsistent approach based on the time dependent Ginzburg Landau (TDGL) theory and taking into account the superconducting fluctuations which cannot be neglected in Bechgaard salts as argued experimentally \cite{JooPhD,Joo04}. This issue, which is the keystone of the present work, has not been addressed so far. The results deduced from our model are consistent with recent experimental data.\newline
We assume that the (TMTSF)$_2$ClO$_4$ sample is formed by ordered domains with a SC phase, embedded in a matrix of insulating islands where 
ClO$_4$ anions are disordered. 
We take for simplicity a layered structure where in each layer, corresponding to the most conducting ($a,b$) plane, we consider an array of identical ordered domains separated by disordered ones which are also assumed to be identical.
The tunneling between the ordered superconducting domains is insured by Josephson couplings denoted $J_1$ and $J_2$ in the $a$ and the $b$ direction respectively while the interplane Josephson coupling is parametrized by $J_0$.\newline
The average size of the SC island is assumed to be greater than the in-plane coherence length $\xi_0=\xi_{ab}= \sqrt{\xi_a\xi_b}\sim 500\AA$ \cite{Pouget}.\\

To bring out the importance of the SC fluctuations in (TMTSF)$_2$ClO$_4$, we evaluate the Ginzburg number $G_i$ which is the fundamental parameter governing the strength of the fluctuations. $G_i$, indicating the temperature range around the critical temperature $T_c$ where fluctuations are not negligible, is given by:
\begin{eqnarray*}
G_i=\frac{\delta T}{T_c} =\frac 1 2 \left[ \frac
{8\pi^2 k_B T_c \lambda^2 \gamma}{ {\phi_0}^2 \xi_{ab} }
\right]^2,
\end{eqnarray*}
here $\gamma=\frac{\xi_{ab}}{\xi_{c}}$ is the anisotropy parameter where $\xi_{ab}$ and $\xi_{c}$ are the coherence lengths in the ($a,b$) plane and along the $c$ axis respectively. $\phi_0$ is the quantum flux and $\lambda$ is the penetration depth.\newline 
Assuming $\xi_{ab}=\sim 500 \AA$, $\gamma\sim 20$ and $\lambda=17 \mu m$ \cite{Revue}, we found $G_i\sim 0.1$, which is considerably large as in high-$T_c$. However, in conventional superconductors the Ginzburg number is small ($G_i\sim 10^{-8}$).
The  large value of $G_i$ in (TMTSF)$_2$ClO$_4$ means that the SC fluctuations spread over a wide interval of temperature compared to the critical temperature $T_c$, which is consistent with the departure from a Fermi liquid behavior of the temperature dependence of the $c$ axis resistivity measured around 5 K and down to $T_c\sim 1.2$ K \cite{JooPhD,Joo04}.\\

Due to the superconducting fluctuations in the granular structure considered in our model, we expect deviations from the mean field theory value $T_0$ corresponding to the transition from the metallic state to the pure superconducting phase where all anions are ordered. We will take $T_0=T_{exp}+Gi\,T_{exp}\sim 1.8$ K where $T_{exp}=1.5$ K is the experimental SC transition temperature found recently in a very slowly cooled pure sample \cite{Shingo,JeromeP}.\\

To study the stability of the SC phase, we compare its Ginzburg-Landau free energy $F_s$ to that of the normal state $F_{norm}$ \cite{Puica}:
\begin{eqnarray}
F&=&F_s-F_{norm}\nonumber\\
&=&\sum_{i,j,n}\int_0^{l_1}dX\int_0^{l_2}dY \left[a|\psi_{nij}|^2+\frac{{\hbar}^2}{2m}|\vec{\nabla}\psi_{nij}|^2
\right.\nonumber\\
&+&\left.J_1|\psi_{nij}-\psi_{n\,i+1\,j}|^2
+J_2|\psi_{nij}-\psi_{n\,i\,j+1}|^2\right.\nonumber\\
&+&\left.J_0|\psi_{nij}-\psi_{n+1\,i\,j}|^2
+\frac b 2 |\psi_{nij}|^4\right]
\label{free}
\end{eqnarray}
where $X$ ($Y$) is the intra-domain coordinate along the $a$ ($b$) axis while $l_1$ and $l_2$ are respectively the lengths of the SC domain along the $a$ and the $b$ directions. $\psi_{nij}$ is the SC order parameter in the $n^{th}$ plane and in the ($i,j$) domain where $i$ ($j$) labels the SC island in the $a$ ($b$) direction. The $a$ coefficient is given by: $a=a_0\epsilon$ where $a_0={\hbar}^2/2m\xi^2_0$ and $\epsilon=\ln(T/T_0)$  while $b=\mu_0\kappa^2e_0^2{\hbar}^2/2m^2$, $\kappa$ is the GL parameter, $e_0=2e$ is the pair electric charge and $m$ is the effective pair mass in the ($ab$) plane.\newline
The coupling parameters are written as:
\begin{equation}
J_1=\frac{{\hbar}^2}{2m^{\ast}l^{{\prime}2}_1}\quad, \quad
J_2=\frac{{\hbar}^2}{2m^{\ast}l^{{\prime}2}_2}\quad, \quad {\mathrm and}
\quad J_0=\frac{{\hbar}^2}{2m_c s^2} 
\label{J}
\end{equation}
with $m^{\ast}$ and $m_c$ are the effective pair masses in the SC domain and along the $c$ axis. We take $m^{\ast}=m$ and we introduce the anisotropic parameter $\gamma=\xi_0/\xi_c=\sqrt{m_c/m}$, where $\xi_c$ is the coherence length is the $c$ direction. $l^{\prime}_1$  and $l^{\prime}_2$ are respectively the size of the insulating domains in the $a$ and $b$ directions whereas $s$ is the interplane distance\

The dynamics of the SC order parameter satisfies the time dependent Ginzburg-Landau (TDGL) equation:
\begin{eqnarray*}
\Gamma^{-1}_0\frac{\partial \psi_{nij}}{\partial t}=-
\frac{\partial F}{\partial\psi^{\ast}_{nij}}+\zeta_{nij}(\vec{r},t)
\end{eqnarray*}
Here $\Gamma^{-1}_0=\pi{\hbar}^3/16 m \xi^2_0 k_BT$ is the relaxation rate of the order parameter whereas $\zeta_{nij}(\vec{r},t)$ are the Langevin forces governing the thermodynamical fluctuations and which fulfill the Gaussian white-noise law\cite{Puica}:
\begin{eqnarray*}
\langle \zeta_{nij}(\vec{r},t) \zeta^{\ast}_{n^{\prime}i^{\prime}j^{\prime}}(\vec{r}\;^{\prime},t^{\prime})
\rangle=2
\Gamma^{-1}_0k_BT\delta(\vec{r}-\vec{r}\;^{\prime})\delta(t-t^{\prime})\frac{\delta_{nn^{\prime}}}s
\end{eqnarray*}
with $\vec{r}=(X+i(l_1+l^{\prime}_1),Y+j(l_2+l^{\prime}_2),ns)$
and $\vec{r}\;^{\prime}=X+i^{\prime}(l_1+l^{\prime}_1),Y+j^{\prime}(l_2+l^{\prime}_2),n^{\prime}s)$.\\

Taking the derivative of the free energy $F$ (Eq.\ref{free}) with respect to $\psi^{\ast}_{nij}$ and replacing $b|\psi_{nij}|^2\psi_{nij}$ by $b\langle|\psi_{nij}|^2\rangle \psi_{nij}$ to get a linear problem, the TDGL equation writes as:
\begin{eqnarray} 
&&\zeta_{nij}(\vec{r},t)=\Gamma^{-1}_0\frac{\partial \psi_{nij}}{\partial t}-\frac{{\hbar}^2}{2m}\Delta\psi_{nij}+a\psi_{nij}\rangle \psi_{nij}\nonumber\\
&+&b\langle|\psi_{nij}|^2\rangle \psi_{nij}+J_1\left(2\psi_{nij}-\psi_{n\,i+1\,j}-\psi_{n\,i-1\,j}\right)\nonumber\\
&+&J_2\left(2\psi_{nij}-\psi_{n\,i\,j+1}-\psi_{n\,i\,j-1}\right)\nonumber\\
&+&J_0\left(2\psi_{nij}-\psi_{n+1\,i\,j}-\psi_{n-1\,i\,j}\right)
\label{zeta}
\end{eqnarray}
We normalize the reduced temperature $\epsilon$:
\begin{eqnarray} 
\tilde{\epsilon}=\epsilon+\frac b a \langle|\psi_{nij}|^2\rangle
\label{self}
\end{eqnarray}
$\tilde{\epsilon}$ and $\langle|\psi_{nij}|^2\rangle$ will be determined self-consistently. We also define the Fourier transform of $\psi_{nij}$ as:
\begin{eqnarray*}
\psi_{nij}(x,y,t)&=&\int\frac{d^2\vec{k}}{(2\pi)^2}\int_{-\frac \pi s}^{\frac \pi s} \frac{dq}{2\pi}\psi_q(k_x,k_y,q,t)\nonumber\\
&&\times{\rm e}^{-ik_x x}\;{\rm e}^{-ik_y y}\;{\rm e}^{-iqns}
\end{eqnarray*}
where
\begin{eqnarray*}
\psi_q(k_x,k_y,t)&=&\sum_{nij}\int_0^{l_1}dX\int_0^{l_2}dY s \nonumber\\
&&\psi_{nij}(X+i(l_1+l^{\prime}_1),Y+j(l_2+l^{\prime}_2),ns,t)\nonumber\\
&&\times{\rm e}^{(X+i(l_1+l^{\prime}_1)k_x}\;
{\rm e}^{(Y+j(l_2+l^{\prime}_2)k_y}\;
{\rm e}^{iqns}
\end{eqnarray*}
Equation \ref{zeta} takes then the form:
\begin{eqnarray}
\zeta_q(k_x,k_y,t)&=&\left[\Gamma^{-1}_0\frac{\partial }{\partial t}
+\frac{{\hbar}^2k^2_x}{2m}+\frac{{\hbar}^2k^2_y}{2m}+\tilde{a}\right.\nonumber\\
&+&2J_1(1-\cos(k_x(l_1+l^{\prime}_1))\nonumber\\
&+&2J_2(1-\cos(k_y(l_2+l^{\prime}_2))\nonumber\\
&+&\left.2J_0(1-\cos(qs)\right]\psi_q(k_x,k_y,q,t)
\label{Fzeta}
\end{eqnarray}
with
\begin{eqnarray*}
\langle\zeta_q(\vec{k},t)\zeta^{\ast}_{q^{\prime}}(\vec{k}^{\prime},t^{\prime})\rangle&=2&\Gamma^{-1}_0 k_bT(2\pi)^3\delta(\vec{k}-\vec{k}^{\prime})
\delta(q-q^{\prime})\nonumber\\
&\times&\delta(t-t^{\prime})
\end{eqnarray*}
where $\vec{k}=(k_x,k_y)$ and $\tilde{a}=a+b\langle|\psi_{nij}|^2\rangle$.

Using the Green function method \cite{Puica}, one can solve Eq.\ref{Fzeta} and determine $\langle|\psi_{nij}|^2\rangle$ by straightforward calculations \cite{else}.\newline
The self-consistent equation Eq.\ref{self} can now be solved to derive the critical temperature T$_c$ at which $\tilde{\epsilon}=0$.\\

In figure 1 we have depicted the superconducting transition temperature as a function of the inverse of the reduced coupling $J^{\prime}_1=J_1/a_0$ along the $a$ axis, which is expected to decrease by quenching the sample since the distances, $l^{\prime}_1$  and $l^{\prime}_2$, separating two SC domains increase with cooling rate, putting at disadvantage the Josephson tunneling. Therefore the $1/J^{\prime}_1$ parameter mimics the cooling rate.
The exact dependence of the Josephson coupling on the cooling rate may be deduced from the behavior of the inter-domain distances as given by Eq.\ref{J}. However, such behavior has not been determined experimentally since X ray diffraction measurements reported only the size parameters of the ordered domains ($l_1$ and $l_2$) which are found to be reduced by quenching \cite{Pouget}. We, then, assume that $l^{\prime}_1$ and $l^{\prime}_2$ increase with the cooling rate and as a consequence $1/J^{\prime}_1$ will be enhanced by quenching. \
\vspace{0.7cm}

\begin{figure}[htpb]
\centerline{\includegraphics[width=7cm,height=6cm]{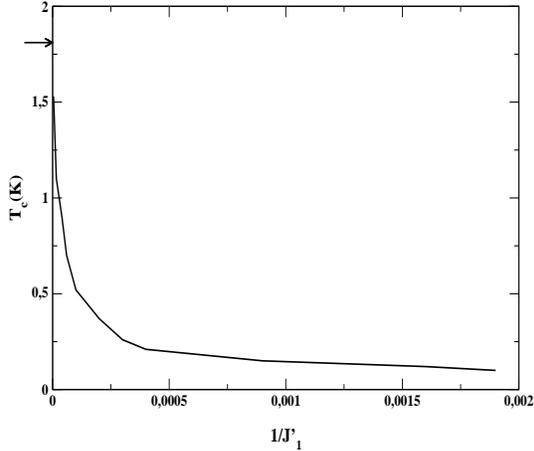}}
\caption{Superconducting transition temperature of the (TMTSF)$_2$ClO$_4$ as a function of the inverse of the reduced Josephson coupling along the $a$ axis which mimics the cooling rate. Numerical results are obtained for $l_1=l_2=800 \AA$, where $l_1$ and $l_2$ are the lengths of the SC domain along the $a$ and the $b$ directions respectively.
The arrow indicates the mean field theory value of the SC transition temperature $T_0$.
}
\label{} 
\end{figure}

As shown in Fig.1, T$_c$, which is lower than the mean field theory value T$_0$ due the SC fluctuations, decreases rapidly for small cooling rates for which the SDW domains can be regarded as point defects having drastic effect on a non-s wave SC phase.
For larger cooling rates, T$_c$ decreases smoothly and tends to saturate with further quenching before ending at a critical value of the Josephson coupling $J^{\prime}_1$ \cite{JooPhD,Mat}. 
It is worth to stress that, actually, T$_c$ does not saturate and continue to decrease but with a very small rate.
However, the greater the cooling rate, the larger the SDW domains, the smaller the anion gap opening on the Fermi surface. The system, which is characterized by a four sheet Fermi surface in the relaxed state, is then more and more close to a one band Fermi surface as in (TMTSF)$_2$PF$_6$.
Therefore, the nesting properties are improved and the transition temperature from the metallic state the SDW phase will then increase by quenching \cite{Heritier84}.\

We have looked for the effect of the SC domain size on T$_c$ since the domains are expected to shrink by quenching\cite{Pouget}. However, we found that T$_c$ is almost independent of the size parameters $l_1$ and $l_2$ of the ordered domains. This is in agreement with our assumption concerning the in-plane coherence length $\xi_0$ which is supposed to remain smaller than the size parameters of the SC grains\cite{JeromeP}.\newline
A point of interest which should be addressed is the striking feature of the upper critical fields of the (TMTSF)$_2$ClO$_4$ measured along the $a$ and the $b$ directions which are found to be greatly enhanced compared to the Pauli limit with an unusual upward curvature \cite{Brown}.
One may expect that the presence of SC and SDW domains in the (TMTSF)$_2$ClO$_4$ may act on the value of the critical fields as found in (TMTSF)$_2$PF$_6$ characterized by the presence of SC slabs sandwiched by SDW phases \cite{Lee}. A simple model based on this slab structure, which has been first proposed by Vuletic {\it et al.}\cite{Vuletic}, gives rise to an enhancement and an upward curvature of the upper critical fields of (TMTSF)$_2$PF$_6$ \cite{Lee}. 
However, we do not claim that the phase segregation picture proposed in the present work may explain the exotic behavior of the critical fields.

\section{Conclusion}

We have investigated, in this letter, the effect of the anion ordering on the SC transition temperature T$_c$ in the (TMTSF)$_2$ClO$_4$ salt taking into account the superconducting fluctuations. The main result of this work is the presence of two regimes in the behavior of T$_c$ on quenching: a first regime with a rapid decrease for small cooling rates where SDW domains act as defects and a second regime with a smooth decrease ending by a saturation before the collapse of the SC phase at large cooling rates for which the SDW domains take over the SC ones.
The obtained results are in good agreement with experimental data. \newline
For a more accurate model, one may consider a random distribution of the SDW and SC islands. However, the general behavior obtained in the present work will be kept unchanged.\newline
Moreover, a complete description of the effect of the cooling rate on the SC phase should include the case where the size of the SC domains is smaller than the coherence length $\xi_0$ as in granular superconductors. This peculiar issue goes beyond the scope of the present work.\newline
However, it is worth to stress that the phase segregation scenario, proposed in this letter, should be borne in mind to account for the striking features of the phase diagram of (TMTSF)$_2$ClO$_4$.
\section{Acknowledgment}
It is a pleasure to warmly thank Pr. M. H\'eritier for fruitful comments and suggestions.
We would like to acknowledge Pr. D. J\'erome, Pr. H. Raffy, Dr. N. Joo, Dr. N. Matsunaga, Dr. L. Fruchter and Pr. J. P. Pouget for helpful and stimulating discussions.
We are indebted to Pr. D. J\'erome and Dr. N. Joo for providing us with their unpublished data.
S. Haddad  and I. Sfar warmly thank the staff of Laboratoire de
Physique des Solides \`a Orsay for kind hospitality.
This work was supported by the French-Tunisian CMCU project 04/G1307.\newline
%
%

\end{document}